\documentclass[runningheads]{llncs}

\usepackage[T1]{fontenc}
\usepackage{graphicx}
\usepackage{amsmath}
\usepackage{booktabs}
\usepackage{array}
\usepackage{url}
\usepackage{float}
\usepackage{marvosym}
\begin{document}

\title{AutoConcept: Training-Free Concept-Guided Reranking for Metadata-Available Composed Image Retrieval}
\titlerunning{AutoConcept for Metadata-Available CIR}

\author{Tianyu Wang\inst{1}\textsuperscript{\Letter} \and Tianjiao Wu\inst{2}}
\authorrunning{T. Wang and T. Wu}
\institute{School of Computer Science and Technology, Soochow University, Suzhou, China\\
\email{tywangcs@foxmail.com}
\and
INSTITUT NATIONAL DES SCIENCES APPLIQUEES DE LYON, Lyon, France\\
\email{tianjiao.wu@insa-lyon.fr}}

\maketitle

\begin{abstract}
Composed image retrieval (CIR) retrieves a target image from a reference image and a text modification. This paper studies metadata-available CIR reranking, where a fixed CIR model first returns a candidate pool and gallery metadata is then used for second-stage concept-guided scoring. We introduce AutoConcept, a training-free reranker that converts concept evidence into an interpretable memory. AutoConcept filters noisy concepts, activates query-relevant positive constraints with an auxiliary negative penalty, and combines base retrieval scores with metadata-based concept-candidate alignment through inference-time calibration. On FashionIQ, AutoConcept yields significant early-rank improvements over WeiMoCIR and consistent plug-in gains on LinCIR candidate pools. Metadata-aware controls show that structured concept memory adds signal beyond direct query-text and extracted-attribute matching, while a query-only variant further supports the effectiveness of concept-level reranking. A supplementary real-human concept-label study indicates that the same memory interface can consume participant-provided evidence. These results position AutoConcept as an interpretable concept-memory reranker for product-style CIR galleries with available metadata.

\end{abstract}

\section{Introduction}

Composed image retrieval (CIR) asks a system to retrieve a target image given a reference image and a natural-language modification. In fashion and e-commerce retrieval, candidate items are commonly associated with metadata such as titles, attributes, tags, or descriptions. Users may also express concept-level constraints, such as preferred colors, sleeve length, neckline, material, or styles to avoid. Most CIR systems construct one composed query representation and rank the gallery by similarity to that query \cite{fashioniq,fashion200k,tirg,val,clip}. A complementary second-stage layer that exposes and applies such concepts can make reranking more controllable and easier to inspect.

This work studies metadata-available concept-guided CIR reranking. A fixed first-stage retriever returns candidate images, and gallery metadata is available only after this candidate pool has been formed. AutoConcept addresses this setting by converting controlled concept evidence into an interpretable memory that guides candidate ordering without training. The central question is whether explicit concept evidence can improve a strong CIR candidate pool beyond direct query-to-metadata or attribute-to-metadata matching.

We propose AutoConcept, a training-free concept-guided reranker. AutoConcept keeps the first-stage CIR model fixed and operates on its top-$K$ candidate pool. It constructs positive concept constraints from controlled evidence, retains a lightweight negative constraint penalty, filters noisy concepts, activates query-relevant concepts with a relevance gate, and combines base retrieval scores with metadata-based concept-candidate alignment. The result is a modular reranking layer that can be attached to different CIR candidate generators.

\begin{figure}[t]
\centering
\includegraphics[width=.80\textwidth]{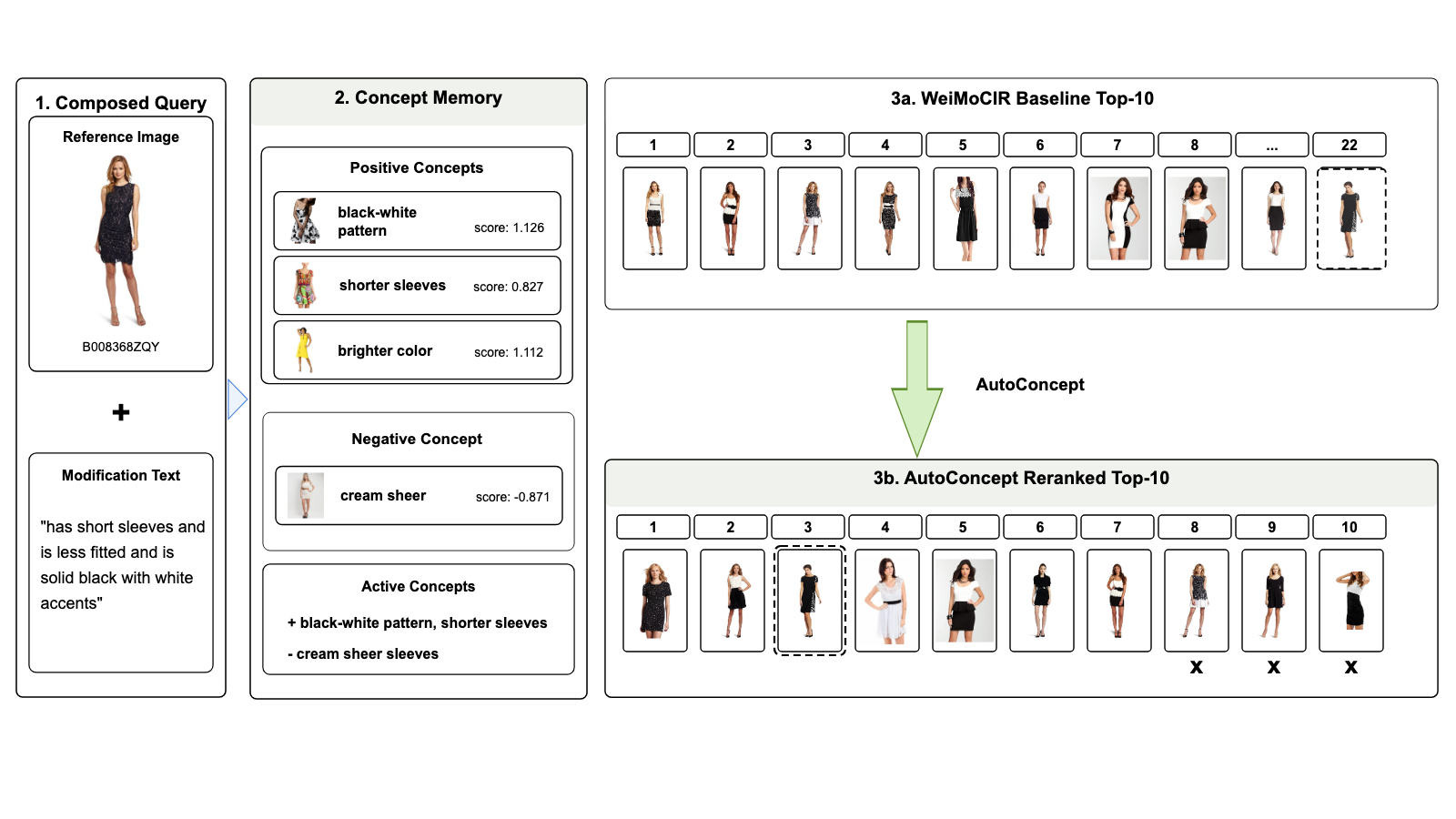}
\caption{AutoConcept in action. Given the same composed query and first-stage candidate pool, the explicit concept memory promotes candidates that match query-relevant positive constraints and penalizes candidates matching negative constraints.}
\label{fig:teaser}
\end{figure}

Our main experiments use FashionIQ with five controlled concept-evidence seeds. AutoConcept improves WeiMoCIR \cite{weimocir} from 0.1125 to 0.1379 Recall@10 and LinCIR candidates \cite{lincir} from 0.2605 to 0.3009 under the same metadata-available reranking protocol. Metadata-aware controls locate the contribution relative to direct query-to-metadata, extracted-attribute, and composed-text reranking; an aligned-vs-shuffled analysis shows that coherent query-evidence correspondence matters. We further collect a real-human concept-label dataset to test whether the same memory interface can consume participant-provided item-level concept evidence.

The contributions are:
\begin{enumerate}
    \item We formulate metadata-available CIR reranking with an explicit concept-memory layer over fixed first-stage candidate pools.
    \item We give an explicit scoring formulation with positive concept constraints, an auxiliary negative penalty, query-concept relevance gating, and closed-form inference-time score calibration.
    \item We evaluate AutoConcept alongside metadata-aware baselines, including query-only evidence, query-to-metadata matching, composed-text metadata matching, and extracted-attribute matching.
    \item We show plug-in gains on both WeiMoCIR and LinCIR candidate pools, with aligned-vs-shuffled concept-evidence analysis and a real-human concept-label study.
\end{enumerate}

\begin{figure}[t]
\centering
\includegraphics[width=.96\textwidth]{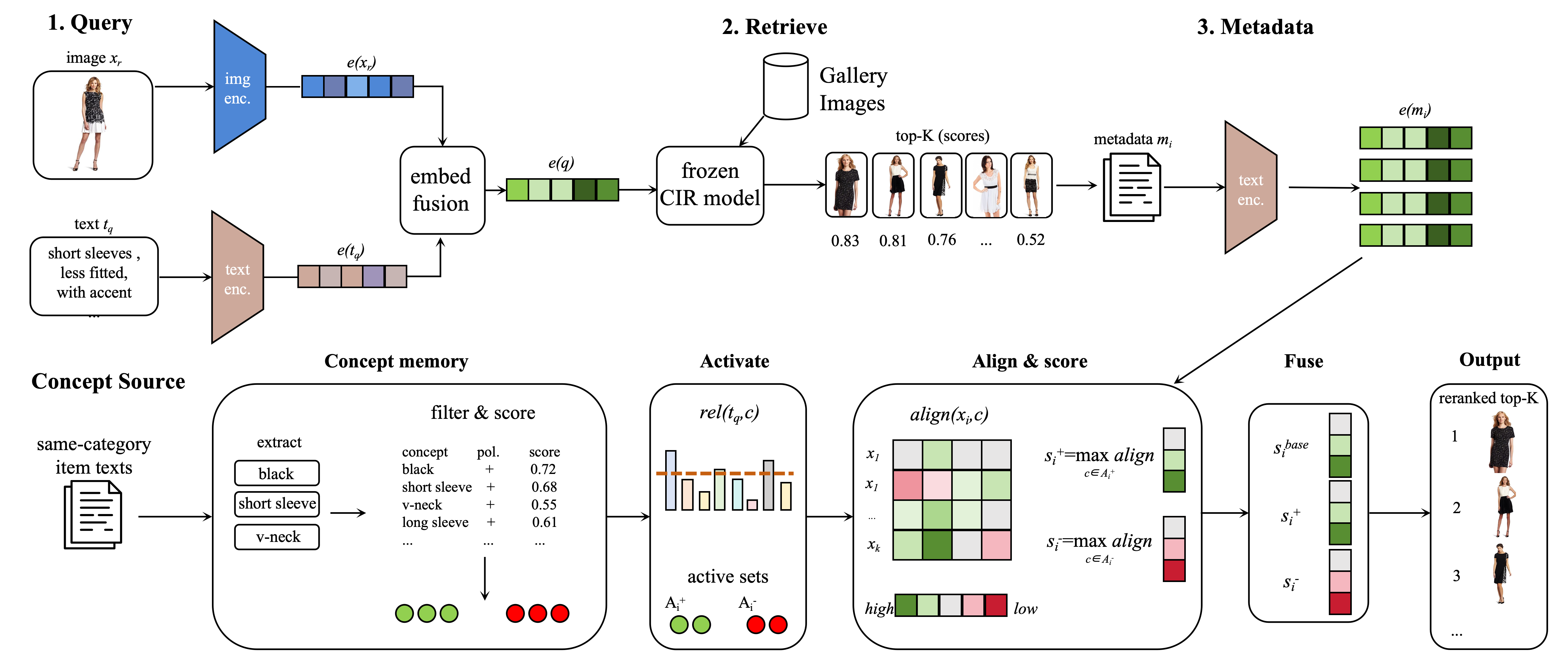}
\caption{Overview of the AutoConcept framework. AutoConcept constructs an explicit concept memory from controlled concept evidence, activates query-relevant concepts, and reranks the top-$K$ candidate pool with inference-time score calibration.}
\label{fig:framework}
\end{figure}

Figure~\ref{fig:framework} summarizes the full AutoConcept pipeline. The base CIR model first produces a top-$K$ candidate pool, while controlled concept evidence is converted into a filtered concept memory. At inference time, query-relevant concepts are activated and used to adaptively rerank the candidate pool.

\section{Related Work}

\subsection{Composed Image Retrieval}

CIR methods retrieve target images using both a reference image and text feedback. Early systems learn image-text composition modules or attention-based fusion over fashion datasets and real-image benchmarks \cite{fashioniq,fashion200k,tirg,val,cirr,searle}. Recent zero-shot, training-free, and language-only CIR systems build on pretrained vision-language embeddings such as CLIP, learn lightweight language-side mappings, or use textual inversion and candidate verification \cite{clip,clip4cir,combiner,pic2word,searle,isearle,lincir,contexti2w,weimocir}. AutoConcept is orthogonal to these retrieval backbones: it adds an explicit concept-memory layer over candidate pools produced by existing CIR methods.

\subsection{Training-Free and Plug-in Reranking}

Zero-shot and low-training CIR systems reduce training cost by weighted modality fusion, score recombination, textual inversion, language-only training, or candidate verification \cite{pic2word,searle,isearle,lincir,contexti2w,weimocir}. This modular view is also related to retrieval-augmented systems, where a frozen generator or scorer is augmented with external evidence \cite{dpr,rag,faiss}. We use WeiMoCIR as the main training-free CIR baseline and keep the same modular design: the base retriever supplies the candidate pool, and concept memory adjusts the order within that pool, making plug-in evaluation possible. Constraint-based rerankers such as SoFT use MLLMs to derive prescriptive and proscriptive constraints \cite{soft}; AutoConcept targets a lighter setting with no per-query MLLM calls, no reranker training, and explicit concept traces.

\subsection{Concept Evidence and Controllable Reranking}

Reranking is a common way to incorporate external evidence after a strong candidate generator, especially when the control signal should remain inspectable. AutoConcept follows this explicit-control direction for CIR and is related to concept-based interpretability methods such as concept activation vectors, automatic concept explanations, concept bottlenecks, prototype networks, basis decomposition, and concept whitening \cite{tcav,ace,cbm}. It is also related to retrieval and recommendation systems that use memory, profiles, or explicit editable attributes as additional evidence \cite{mf,bpr,ncf,memorynet,myvlm,rap}. In AutoConcept, concepts have names, polarity, strength, grounding scores, and evidence items, so the memory can be inspected and diagnosed.

\section{Method}

\subsection{Task Setup}

Each query consists of a reference image $x_r$, modification text $t_q$, and a target item $y$ in a gallery. A base retrieval model returns a ranked candidate pool $C_q=\{x_i\}_{i=1}^{K}$. In the main FashionIQ experiments, $K=100$. AutoConcept reranks only this candidate pool.

Each query is associated with a controlled concept-evidence set $E_q$. AutoConcept converts this evidence into positive concept constraints $P_q$ and a small auxiliary set of negative concept constraints $N_q$. A concept stores a normalized name, polarity, score, grounding score, and evidence count.

\subsection{Information Source Protocol}

AutoConcept uses information available under the metadata-available reranking protocol. For each evaluation query, concept evidence is constructed before reranking and excludes the ground-truth target identity, target image, target metadata, candidate ranks, and post-retrieval relevance labels. The main controlled evidence samples same-category training-side item texts and converts them into concept constraints with a fixed extraction and filtering pipeline. Candidate metadata is used only at reranking time to score items already returned by the first-stage retriever. We additionally evaluate a query-only evidence variant, where concepts are extracted from the modification text $t_q$.

\begin{table}
\caption{Information sources in the FashionIQ metadata-available protocol.}
\label{tab:evidence_sources}
\centering
\scriptsize
\setlength{\tabcolsep}{3pt}
\begin{tabular}{p{.28\textwidth}p{.10\textwidth}p{.48\textwidth}}
\toprule
Source & Used? & Purpose\\
\midrule
Reference image $x_r$ & Yes & first-stage CIR query\\
Modification text $t_q$ & Yes & query encoding and concept activation\\
Training-side evidence item text & Yes & controlled concept-evidence construction\\
Candidate metadata $m_i$ & Yes & second-stage candidate-concept alignment\\
Target image / target metadata & No & excluded from evidence construction and scoring\\
Ground-truth target ID & No & excluded from scoring and evidence construction\\
Candidate rank / relevance label & No & excluded from evidence construction\\
\bottomrule
\end{tabular}
\end{table}

\subsection{Concept Memory Construction}

AutoConcept extracts short attribute phrases from concept-evidence item text, embeds phrases with a sentence encoder, clusters near-duplicate phrases, and scores concepts with positive or negative evidence. A compact set of quality-control constraints removes generic names, polarity conflicts, poorly grounded concepts, and weak concepts. The filter keeps frequent but meaningful apparel attributes such as sleeves, collar, buttons, stripes, and V-neck.

\subsection{Query-Relevant Concept Activation}

For a query $q$ and concept $c$, query-concept relevance is
\begin{equation}
r(q,c)=\cos(e(t_q),e(c)),
\end{equation}
where $e(\cdot)$ is the sentence-encoder embedding and $c$ is the concept name. A query-relevance gate activates the concept set:
\begin{equation}
A_q=\{c:r(q,c)\geq \tau\}.
\end{equation}
AutoConcept can also compute a query-specific gate for inference-time calibration:
\begin{equation}
\tau_q=\operatorname{clip}(\mu(R_q)+\kappa\sigma(R_q),\tau_{min},\tau_{max}),
\end{equation}
where $R_q=\{r(q,c):c\in P_q\cup N_q\}$. The clipping range is fixed before evaluation and shared across runs. Appendix~\ref{app:sensitivity} reports sensitivity around the operating region.

\subsection{Candidate-Concept Alignment}

For active concept $c$ and candidate item $x_i$, AutoConcept computes
\begin{equation}
\operatorname{align}(x_i,c)=\cos(e(m_i),e(c)),
\end{equation}
where $m_i$ is candidate item metadata text. In our experiments, this text comes from the gallery item metadata used uniformly for candidate items; it is not generated from the test target image and does not use the ground-truth target identity during scoring. This defines a metadata-available reranking protocol; Section~\ref{sec:metadata_controls} compares AutoConcept with a direct metadata-text reranker under the same information setting.

Positive concept scores and the auxiliary disliked-concept penalty are aggregated by max pooling:
\begin{equation}
s^+_q(x_i)=\max_{c\in A^+_q}\operatorname{align}(x_i,c), \quad
s^-_q(x_i)=\max_{c\in A^-_q}\operatorname{align}(x_i,c),
\end{equation}
with empty active sets assigned zero contribution. The concept score is
\begin{equation}
s^{concept}_q(x_i)=w_q(q)s^{base}_q(x_i)+w_p(q)s^+_q(x_i)-w_n(q)s^-_q(x_i).
\end{equation}
The weights are set at inference time from query-specific concept evidence. Let
\begin{equation}
a^+_q=\max A^+_q,\quad a^-_q=\max A^-_q,\quad a_q=\max(a^+_q,a^-_q),
\end{equation}
where $A^+_q$ and $A^-_q$ are the gated query-concept activation scores, and the maximum over an empty active set is zero. We also compute base-retriever confidence from the top-score margin:
\begin{equation}
b_q=\operatorname{clip}\left(\frac{s^{base}_{(1)}-s^{base}_{(2)}}{\delta},0,1\right).
\end{equation}
Here $s^{base}_{(1)}$ and $s^{base}_{(2)}$ are the largest and second largest base scores in the candidate pool, and $\delta$ is a fixed margin reference shared by all experiments. Positive and negative reliability scales are
\begin{equation}
c^+_q=c_{min}+(1-c_{min})\operatorname{clip}\left(\frac{a^+_q-a_{min}}{a_{max}-a_{min}},0,1\right),
\end{equation}
\begin{equation}
c^-_q=c_{min}+(1-c_{min})\operatorname{clip}\left(\frac{a^-_q-a_{min}}{a_{max}-a_{min}},0,1\right).
\end{equation}
The final per-query weights are
\begin{equation}
w_p(q)=w_p^{cap}c^+_q(1-0.5b_q),\quad
w_n(q)=w_n^{cap}c^-_q(1-0.5b_q),
\end{equation}
\begin{equation}
w_q(q)=w_q^{min}+\max(w_p^{cap}-w_p(q),0)+\max(w_n^{cap}-w_n(q),0).
\end{equation}
The caps are fixed upper bounds, while the actual weights are decided for each query. Thus, when concept evidence is weak or the base retriever is confident, more score mass stays with the base retriever; when active concepts are strong and the base margin is small, positive concepts receive larger weights. The disliked-concept term is intentionally lightweight and acts as a conservative penalty. Image, text, and concept embeddings are L2-normalized before cosine similarity. This closed-form inference-time calibration makes the scoring coefficients query-dependent without training additional parameters.

\subsection{Inference-Time Score Interpolation}

AutoConcept further sets the outer interpolation coefficient from inference-time confidence so weak concept activation does not override the base retriever. Let $\bar a_q$ be the mean activation strength of active concepts for query $q$. The upper interpolation bound and base value are set from the base margin:
\begin{equation}
\lambda^{max}_q=\lambda_{min}+(\lambda^{cap}_{max}-\lambda_{min})(1-b_q),
\end{equation}
\begin{equation}
\lambda^{base}_q=\lambda_{min}+(\lambda_0-\lambda_{min})(1-b_q),
\end{equation}
\begin{equation}
\lambda_q=\operatorname{clip}(\lambda^{base}_q+\gamma \bar a_q,\lambda_{min},\lambda^{max}_q).
\end{equation}
The interpolation bounds are fixed before evaluation and produce a query-specific coefficient. The final score is
\begin{equation}
s_q(x_i)=(1-\lambda_q)s^{base}_q(x_i)+\lambda_qs^{concept}_q(x_i).
\end{equation}

\begin{table}
\caption{AutoConcept reranking procedure.}
\label{tab:algorithm}
\centering
\small
\setlength{\tabcolsep}{5pt}
\begin{tabular}{ll}
\toprule
Step & Operation\\
\midrule
1 & Take query $(x_r,t_q)$, candidate pool $C_q$, metadata $\{m_i\}$, and evidence $E_q$\\
2 & Extract, cluster, score, and filter concept constraints from $E_q$\\
3 & Compute query-concept relevance $r(q,c)$ and activate concepts\\
4 & Align each candidate metadata text $m_i$ with active concepts\\
5 & Compute positive score, auxiliary negative penalty, and base confidence\\
6 & Infer $w_q(q),w_p(q),w_n(q)$ and $\lambda_q$ from query-time signals\\
7 & Rerank $C_q$ by the final score $s_q(x_i)$\\
\bottomrule
\end{tabular}
\end{table}

\section{Experiments}

\subsection{Dataset, Protocol, and Baselines}

The main dataset is FashionIQ. We evaluate five controlled concept-evidence seeds and report mean with sample standard deviation when a method depends on constructed evidence. The first-stage baseline is WeiMoCIR:
\begin{equation}
q=0.20e(x_r)+0.80e(t_q).
\end{equation}
Candidate scoring uses candidate image embeddings for first-stage retrieval. Candidate metadata is introduced only in second-stage reranking controls and AutoConcept. The default candidate pool is top-100. We report Recall@10, Recall@50, MRR, and NDCG@10.

We evaluate metadata-aware rerankers on the same candidate pool: query-text to metadata matching, composed reference-text plus query-text to metadata matching, extracted query-attribute to metadata matching, and a resource-matched prescriptive/proscriptive constraint matcher. These baselines locate AutoConcept relative to direct metadata and constraint matching. The query-only AutoConcept variant builds concepts from $t_q$ and evaluates the concept-memory scorer with query-local evidence.

\begin{table}
\caption{Main results on two first-stage retrievers. AutoConcept reranks the fixed candidate pool under the same metadata-available protocol.}
\label{tab:main}
\centering
\scriptsize
\setlength{\tabcolsep}{2pt}
\begin{tabular}{llrrrr}
\toprule
First-stage & Reranker & R@10 & R@50 & MRR & NDCG@10\\
\midrule
WeiMoCIR & none & 0.1125 & 0.2256 & 0.0604 & 0.0680\\
WeiMoCIR & AutoConcept & 0.1379 $\pm$ 0.0011 & 0.2452 $\pm$ 0.0018 & 0.0739 $\pm$ 0.0008 & 0.0844 $\pm$ 0.0009\\
LinCIR & none & 0.2605 & 0.4618 & 0.1449 & 0.1637\\
LinCIR & AutoConcept & 0.3009 $\pm$ 0.0018 & 0.4915 $\pm$ 0.0023 & 0.1677 $\pm$ 0.0010 & 0.1912 $\pm$ 0.0008\\
\bottomrule
\end{tabular}
\end{table}

Table~\ref{tab:main} shows consistent gains on both candidate generators, concentrated in early-rank metrics.

\begin{figure}[t]
\centering
\includegraphics[width=.82\textwidth]{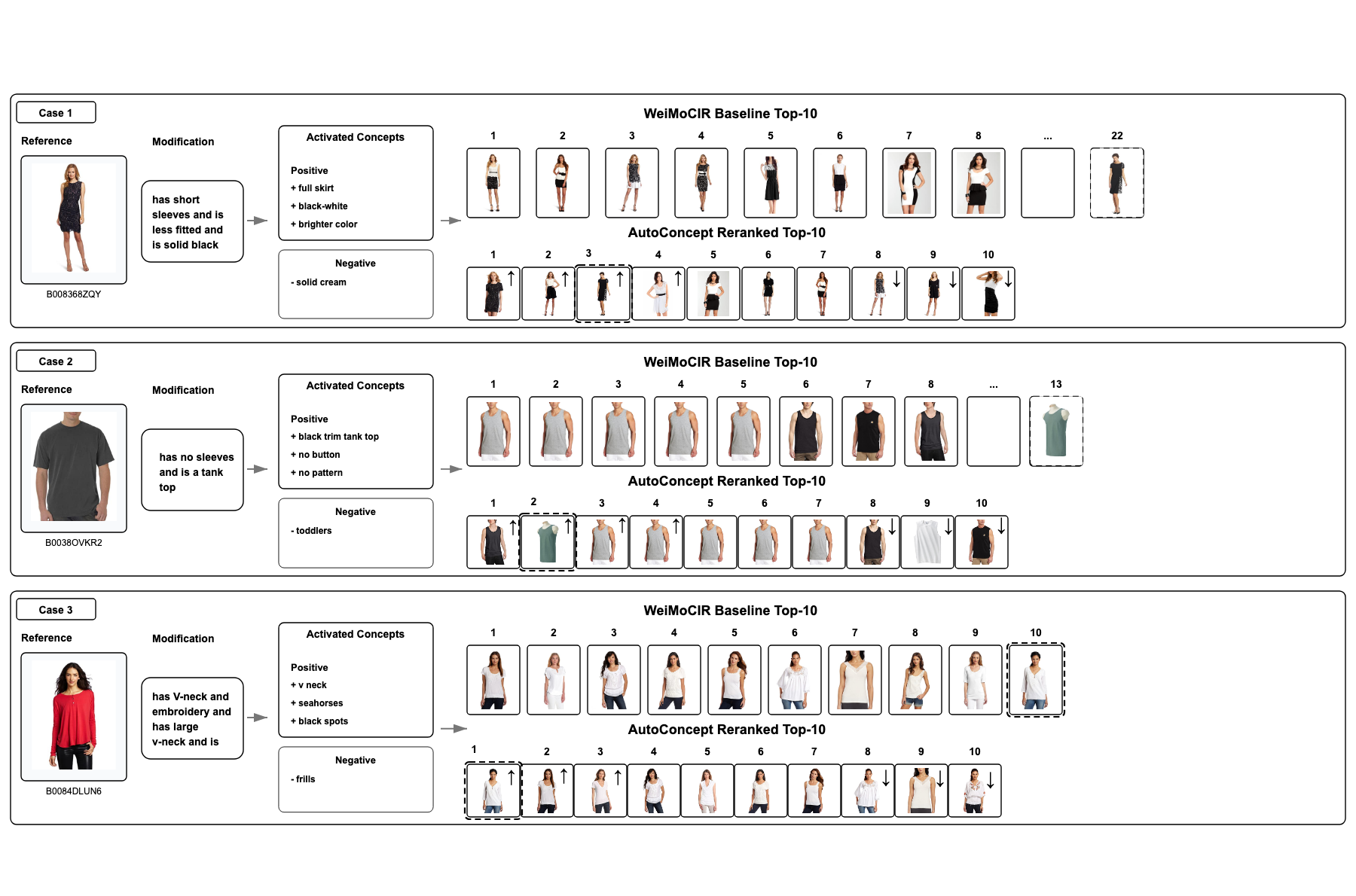}
\caption{Qualitative reranking examples. AutoConcept uses active concepts to adjust the top-10 candidate list from WeiMoCIR, improving concept-evidence alignment while retaining the fixed first-stage candidate pool.}
\label{fig:qualitative}
\end{figure}

\subsection{Significance Testing}

We use query-level paired bootstrap tests. Each FashionIQ query is one resampling unit; AutoConcept scores are first averaged over five concept-evidence seeds and then paired with the same query under the base retriever. We resample queries with replacement to compute percentile 95\% confidence intervals and two-sided paired bootstrap $p$-values. The reported $p$-values are descriptive and are not adjusted for multiple metrics; seed-level robustness is reported through the standard deviations in Table~\ref{tab:main}.

\begin{table}
\caption{Query-level paired bootstrap significance against WeiMoCIR.}
\label{tab:sig}
\centering
\small
\setlength{\tabcolsep}{4pt}
\begin{tabular}{lrrrl}
\toprule
Metric & Delta & Relative & 95\% CI & $p$-value\\
\midrule
R@10 & 0.0254 & +22.54\% & [0.0223, 0.0285] & $<0.001$\\
R@50 & 0.0196 & +8.71\% & [0.0163, 0.0230] & $<0.001$\\
MRR & 0.0135 & +22.26\% & [0.0118, 0.0151] & $<0.001$\\
NDCG@10 & 0.0164 & +24.16\% & [0.0147, 0.0182] & $<0.001$\\
\bottomrule
\end{tabular}
\end{table}

The LinCIR deltas in Table~\ref{tab:main} are also significant with $p<0.001$ under the same query-level paired bootstrap protocol.

\subsection{Metadata-Aware Reranking Controls}
\label{sec:metadata_controls}

Table~\ref{tab:metadata_controls} evaluates AutoConcept alongside metadata-aware rerankers on the same top-100 WeiMoCIR pool. Query-only AutoConcept improves over query-only text-to-metadata, extracted-attribute reranking, and resource-matched constraint matching on early-rank metrics. The reference-text enriched row concatenates reference-item metadata with the modification text and is reported as a strong text-only upper reference under a richer textual signal rather than a same-evidence concept-memory baseline.

\begin{table}
\caption{Metadata-aware reranking controls on FashionIQ seed 1. Query-only AutoConcept uses concepts extracted only from the modification text.}
\label{tab:metadata_controls}
\centering
\scriptsize
\setlength{\tabcolsep}{2pt}
\begin{tabular}{lccrrrr}
\toprule
Setting & Metadata & Memory & R@10 & R@50 & MRR & NDCG@10\\
\midrule
WeiMoCIR & No & No & 0.1125 & 0.2256 & 0.0604 & 0.0680\\
Query text $\rightarrow$ metadata & Yes & No & 0.1318 & 0.2513 & 0.0731 & 0.0822\\
Extracted attributes $\rightarrow$ metadata & Yes & No & 0.1363 & 0.2591 & 0.0754 & 0.0850\\
Constraint matching & Yes & No & 0.1383 & 0.2603 & 0.0763 & 0.0861\\
Query-only AutoConcept & Yes & Yes & 0.1400 & 0.2630 & 0.0800 & 0.0893\\
Controlled-evidence AutoConcept & Yes & Yes & 0.1375 & 0.2435 & 0.0741 & 0.0845\\
\midrule
Reference text + query $\rightarrow$ metadata & Yes & No & 0.1430 & 0.2645 & 0.0811 & 0.0910\\
\bottomrule
\end{tabular}
\end{table}

\subsection{Concept-Evidence Alignment}

Table~\ref{tab:main} is the main controlled concept-evidence setting. Separately, to test whether AutoConcept uses query-relevant concept evidence rather than only category priors, we compare aligned evidence with shuffled evidence. The shuffled control preserves the same domain and category-level concept distribution but breaks query-evidence alignment.

\begin{table}
\caption{Concept-evidence alignment analysis. This setting is separate from the main controlled concept-evidence setting in Table~\ref{tab:main}.}
\label{tab:evidence_alignment}
\centering
\scriptsize
\setlength{\tabcolsep}{2pt}
\begin{tabular}{lrrrrr}
\toprule
Variant & Base R@10 & R@10 & R@50 & MRR & NDCG@10\\
\midrule
Aligned evidence & 0.1125 & 0.1892 $\pm$ 0.0028 & 0.2715 $\pm$ 0.0014 & 0.1050 $\pm$ 0.0010 & 0.1213 $\pm$ 0.0014\\
Shuffled evidence & 0.1125 & 0.1388 $\pm$ 0.0010 & 0.2469 $\pm$ 0.0010 & 0.0738 $\pm$ 0.0008 & 0.0845 $\pm$ 0.0004\\
Delta & -- & +0.0504 & +0.0246 & +0.0312 & +0.0368\\
\bottomrule
\end{tabular}
\end{table}

The aligned setting is substantially stronger than the shuffled control, supporting the role of coherent query-evidence alignment.

\subsection{Real-Human Concept Labels}

We also evaluate AutoConcept with a pilot real-human concept-label split. Participants labeled FashionIQ gallery items and selected optional item-level concept tags. Many labeled items have empty or very short original item text, so the selected tags provide external concept evidence beyond rich product metadata. We build participant-level concept memories from the deduplicated labels and rerank the same top-100 WeiMoCIR pool.

Table~\ref{tab:human_real} evaluates whether human item-level judgments can be converted into useful concept memory under the same metadata-available reranking pipeline.

\begin{table}
\caption{Real-human concept-label evaluation on FashionIQ.}
\label{tab:human_real}
\centering
\scriptsize
\setlength{\tabcolsep}{3pt}
\begin{tabular}{lrrrr}
\toprule
Setting & R@10 & R@50 & MRR & NDCG@10\\
\midrule
WeiMoCIR candidate pool & 0.1125 & 0.2256 & 0.0604 & 0.0680\\
WeiMoCIR + human AutoConcept & 0.1159 & 0.2311 & 0.0624 & 0.0702\\
Delta & +0.0034 & +0.0056 & +0.0019 & +0.0022\\
\bottomrule
\end{tabular}
\end{table}

The participant histories are sparse and are not collected for specific benchmark targets, yet all four target-label metrics improve, showing that AutoConcept can ingest human concept labels under annotation noise. The smaller gain compared with controlled evidence is expected: participants label only a limited subset of items, selected tags may not cover every FashionIQ modifier, and human memories are not constructed around the benchmark target for each query.

\subsection{Ablation Analysis}

Table~\ref{tab:components} isolates scoring components on the same FashionIQ seed-1 top-100 pool. Max pooling is important because concept matches are sparse. Removing the lightweight negative penalty changes little, confirming that FashionIQ gains are driven mainly by positive concept activation.

\begin{table}
\caption{Ablation analysis on FashionIQ seed 1.}
\label{tab:components}
\centering
\small
\setlength{\tabcolsep}{4pt}
\begin{tabular}{lrrr}
\toprule
Setting & R@10 & MRR & NDCG@10\\
\midrule
WeiMoCIR & 0.1125 & 0.0604 & 0.0680\\
Full AutoConcept & 0.1375 & 0.0741 & 0.0845\\
Positive concepts only & 0.1380 & 0.0742 & 0.0847\\
Stronger negative penalty & 0.1363 & 0.0738 & 0.0840\\
Mean pooling & 0.1240 & 0.0679 & 0.0764\\
Top-$k$ mean pooling & 0.1270 & 0.0695 & 0.0783\\
Dynamic gate without outer interpolation & 0.1302 & 0.0703 & 0.0798\\
\bottomrule
\end{tabular}
\end{table}

\section{Scope and Extensions}

AutoConcept is scoped as a second-stage reranker for metadata-available product-style galleries. The main experiments use controlled concept evidence, while real-human annotations provide participant-supplied evidence under natural label noise and sparse item text. The method is most suitable when gallery metadata is available and reasonably aligned with item appearance; missing or inconsistent metadata remains a practical boundary. Broader domains, richer visual grounding, larger human-preference studies, stronger metadata denoising, and heavier learned verification rerankers are natural extensions under different compute budgets.

\section{Conclusion}

AutoConcept adds an explicit concept-memory layer to metadata-available composed image retrieval, complementing retrieval-augmented and concept-controlled multimodal systems \cite{blip2,llava,rap}. It improves both WeiMoCIR and LinCIR candidate pools, compares favorably with direct query and attribute metadata matching, and provides an inspectable memory mechanism. The real-human concept-label study further shows that the same interface can consume participant-provided evidence.

\appendix

\section{Additional Evaluation Results}

\subsection{Gate Sensitivity}
\label{app:sensitivity}

Table~\ref{tab:gate_fiq} reports a gate-sensitivity analysis around $\tau=0.35$. It changes only the query-concept relevance gate and keeps the top-100 candidate pool, controlled concept-evidence setting, concept memory, and calibration bounds fixed. The current implementation can infer $\tau_q$ per query from the memory relevance distribution.

\begin{figure}[t]
\centering
\includegraphics[width=.84\textwidth]{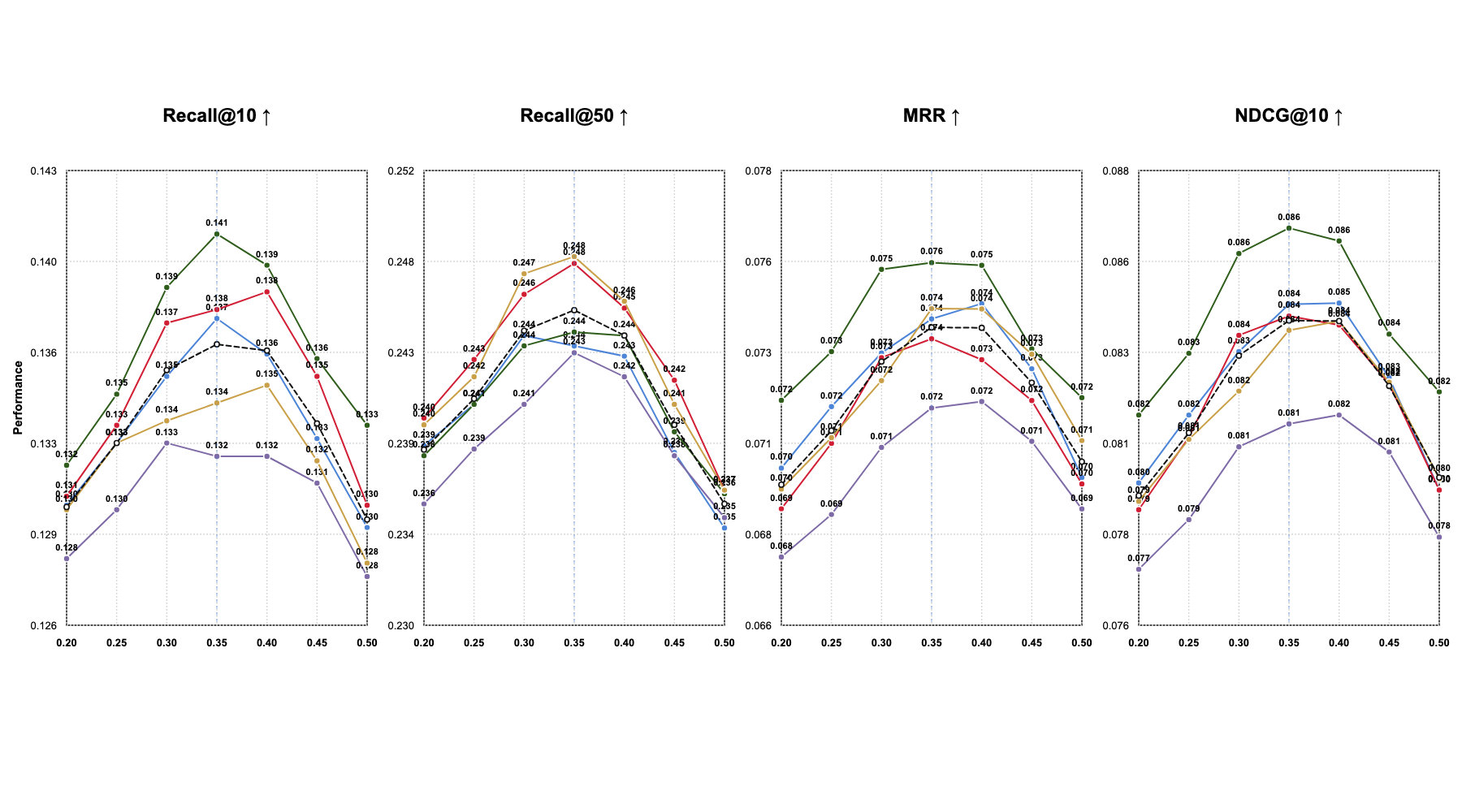}
\caption{Gate sensitivity on FashionIQ. AutoConcept remains stable around the operating range used by inference-time gate calibration.}
\label{fig:gate_sensitivity}
\end{figure}

\begin{table}
\caption{FashionIQ gate sensitivity.}
\label{tab:gate_fiq}
\centering
\scriptsize
\setlength{\tabcolsep}{2pt}
\begin{tabular}{lrrrr}
\toprule
Gate & R@10 & R@50 & MRR & NDCG@10\\
\midrule
0.20 & 0.1304 $\pm$ 0.0013 & 0.2385 $\pm$ 0.0016 & 0.0697 $\pm$ 0.0015 & 0.0794 $\pm$ 0.0015\\
0.25 & 0.1328 $\pm$ 0.0016 & 0.2410 $\pm$ 0.0016 & 0.0711 $\pm$ 0.0016 & 0.0811 $\pm$ 0.0016\\
0.30 & 0.1355 $\pm$ 0.0024 & 0.2442 $\pm$ 0.0024 & 0.0730 $\pm$ 0.0017 & 0.0831 $\pm$ 0.0019\\
0.35 & 0.1379 $\pm$ 0.0011 & 0.2452 $\pm$ 0.0018 & 0.0739 $\pm$ 0.0008 & 0.0844 $\pm$ 0.0009\\
0.40 & 0.1363 $\pm$ 0.0028 & 0.2440 $\pm$ 0.0015 & 0.0738 $\pm$ 0.0014 & 0.0840 $\pm$ 0.0016\\
0.45 & 0.1335 $\pm$ 0.0020 & 0.2397 $\pm$ 0.0016 & 0.0724 $\pm$ 0.0010 & 0.0823 $\pm$ 0.0011\\
0.50 & 0.1300 $\pm$ 0.0022 & 0.2359 $\pm$ 0.0009 & 0.0703 $\pm$ 0.0011 & 0.0799 $\pm$ 0.0014\\
\bottomrule
\end{tabular}
\end{table}

The best region is broad. Gates from 0.30 to 0.40 yield similar gains, and all tested values remain above WeiMoCIR.
Positive concepts drive most of the observed gain in the current FashionIQ setting. Disliked concepts are retained as a lightweight, inspectable penalty channel rather than as the primary source of improvement.

\subsection{Candidate Pool Size}

Table~\ref{tab:topk} varies the first-stage candidate pool size for seed 1. The default top-100 pool gives the strongest AutoConcept result in this analysis. Top-50 limits candidate coverage, while larger pools introduce more metadata noise for the same concept-memory scorer.

\begin{table}
\caption{Candidate-pool sensitivity on FashionIQ seed 1.}
\label{tab:topk}
\centering
\small
\setlength{\tabcolsep}{4pt}
\begin{tabular}{rrrrr}
\toprule
$K$ & Base R@10 & AC R@10 & Base MRR & AC MRR\\
\midrule
50 & 0.1125 & 0.1144 & 0.0594 & 0.0609\\
100 & 0.1125 & 0.1375 & 0.0604 & 0.0741\\
200 & 0.1127 & 0.1165 & 0.0608 & 0.0631\\
500 & 0.1129 & 0.1180 & 0.0614 & 0.0645\\
\bottomrule
\end{tabular}
\end{table}

This sensitivity suggests that AutoConcept is most effective when the first-stage retriever supplies a reasonably precise candidate pool. For weaker retrievers that require very large pools, stronger metadata denoising or visual verification would be needed before reranking.

\subsection{Fashion200K Secondary Check}

Fashion200K provides a secondary fashion-domain check constructed from item text and sampled candidate pools. Table~\ref{tab:fashion200k_secondary} reports three-seed means. AutoConcept improves all four metrics over the same WeiMoCIR candidate generator.

\begin{table}
\caption{Fashion200K secondary retrieval check.}
\label{tab:fashion200k_secondary}
\centering
\scriptsize
\setlength{\tabcolsep}{3pt}
\begin{tabular}{lrrrr}
\toprule
Setting & R@10 & R@50 & MRR & NDCG@10\\
\midrule
WeiMoCIR & 0.6560 $\pm$ 0.0339 & 0.9144 $\pm$ 0.0060 & 0.3458 $\pm$ 0.0097 & 0.4081 $\pm$ 0.0167\\
WeiMoCIR + AutoConcept & 0.7049 $\pm$ 0.0167 & 0.9355 $\pm$ 0.0013 & 0.3709 $\pm$ 0.0203 & 0.4404 $\pm$ 0.0199\\
\bottomrule
\end{tabular}
\end{table}

\subsection{Runtime and Latency}

\begin{table}
\caption{Runtime comparison between the first-stage WeiMoCIR retriever and the AutoConcept second-stage overhead. The AutoConcept loop column isolates the concept-scoring loop after embeddings are available.}
\label{tab:runtime}
\centering
\small
\setlength{\tabcolsep}{4pt}
\begin{tabular}{lrrrrrr}
\toprule
Dataset & Queries & Gallery & Mean $K$ & Base retriever & AC full & AC loop\\
\midrule
FashionIQ & 6016 & 74381 & 100.00 & 5.7290 & 1.8080 & 0.3199\\
Fashion200K & 413 & 40000 & 62.00 & 3.1184 & 6.2910 & 0.1928\\
\bottomrule
\end{tabular}
\end{table}

Table~\ref{tab:runtime} reports milliseconds per query on an Apple M5 MacBook Air with 24GB unified memory, batch size one, PyTorch/MPS where available, and precomputed CLIP image embeddings. In the cached setting, AutoConcept adds 0.3199 ms/query on FashionIQ and 0.1928 ms/query on Fashion200K, corresponding to 5.58\% and 6.18\% of first-stage retrieval latency. The higher Fashion200K AC full cost comes from uncached text-encoder setup and metadata encoding over a smaller query set; these embeddings are precomputable in the reranking use case.

\section{Implementation Details}

\subsection{Candidate Pool and Hyperparameters}

The main FashionIQ experiment reranks the top-100 candidates from the base retriever. The same top-100 pool is used for WeiMoCIR, AutoConcept, and metadata-aware controls. Concept names, query text, and metadata text are encoded with all-MiniLM-L6-v2; CLIP ViT-B/32 is used for image embeddings and CLIP-based grounding. Quality-control rules and calibration bounds are fixed before evaluation and shared across seeds, candidate generators, and metadata-aware controls; no query labels, target ranks, or relevance labels are used for tuning.

\begin{table}
\caption{Reproducibility settings.}
\label{tab:hparams}
\centering
\scriptsize
\setlength{\tabcolsep}{3pt}
\begin{tabular}{ll}
\toprule
Setting & Role\\
\midrule
WeiMoCIR query fusion & fixed weighted fusion of reference image and modification text\\
First-stage candidate scoring & image-side scoring; candidate metadata enters only in reranking\\
Candidate pool & top-100 candidates unless stated otherwise\\
Concept/text encoder & all-MiniLM-L6-v2 for concept names, queries, and metadata\\
Image encoder and grounding & CLIP ViT-B/32 image embeddings and CLIP-based grounding\\
Quality control & fixed filtering rules shared across seeds and candidate generators\\
Score calibration & bounded query-adaptive weights and interpolation, fixed before evaluation\\
Tuning data & no query labels, target ranks, or relevance labels used for tuning\\
\bottomrule
\end{tabular}
\end{table}

For $K$ candidates, embedding dimension $d$, and active concept sets $A^+_q$ and $A^-_q$, the reranking loop scales as $O(K(|A^+_q|+|A^-_q|)d)$. The relevance gate keeps the active concept set small.

\section{Dataset and Concept-Evidence Construction Details}

FashionIQ is the main CIR benchmark. Each query contains a reference image, two modification captions, and a target image. We construct controlled concept-evidence sets from training-side item texts; the query-only control extracts concept names only from the modification text.

Fashion200K is a secondary fashion-domain robustness check with composed queries from item text and sampled candidate pools.

The aligned analysis uses controlled attribute families such as color, printed patterns, sleeve length, neckline, length, and fit. It is separate from the main protocol and excludes the target image, target metadata, candidate ranks, and ground-truth target identity. The shuffled control keeps the evidence distribution but shuffles query-evidence mappings within category.

The real-human study uses a separate collection interface. Participants label FashionIQ items and select optional tags from a fixed vocabulary covering color, pattern, sleeve, neckline, length, fit, and style. Labels are deduplicated, converted into the same concept-memory schema, and stored as a separate human-data split.

\section{Concept Memory and Filtering Details}

\begin{table}
\caption{Main concept filter rules.}
\label{tab:filter_rules}
\centering
\begin{tabular}{lll}
\toprule
Rule & Action & Criterion\\
\midrule
Generic concept name & drop & fixed generic-term stoplist\\
Positive/negative conflict & drop & same normalized name in both memories\\
Low CLIP grounding & drop & below the fixed grounding quality constraint\\
Low concept strength & drop & below the fixed concept-strength constraint\\
Low evidence & drop & evidence count $<1$\\
High global frequency & keep & disabled in main filter\\
Single weak evidence & keep & disabled in main filter\\
Minimum positive memory & restore & at least one positive if available\\
\bottomrule
\end{tabular}
\end{table}

The filter removes obvious generic or weak concepts while keeping visually meaningful apparel attributes such as sleeve length, neckline, buttons, stripes, and V-neck.

\section{Explanation Diagnostics and Ethics}

AutoConcept can list active concepts, evidence items, and candidate-side matches, exposing how memory entries participate in reranking.

AutoConcept stores explicit named concepts rather than opaque user embeddings, supporting inspection, deletion, and editing. The real-human study uses voluntary item-level labels without requesting sensitive user attributes; larger deployments should avoid protected-attribute inference and expose memory-retention controls.

\begin{credits}
\subsubsection{\ackname} We thank the anonymous reviewers for their constructive comments and the participants who contributed item-level concept labels.

\subsubsection{\discintname}
The authors have no competing interests to declare that are relevant to the content of this article.
\end{credits}

\end{document}